\documentclass{elsart}

\usepackage{graphicx}

\usepackage{amssymb}

\begin{document}

\begin{frontmatter}

\title{A simple model for a transverse dune field}

\author{E. J. R. Parteli$^{1}$\corauthref{ejrp}}\corauth[ejrp]{Corresponding author. Tel.: +49-711-685-3593; 
fax: +49-711-685-3658}\ead{parteli@ica1.uni-stuttgart.de}\and\author{H. J. Herrmann$^{1,2}$}
\begin{center}
\address{(1) Institut fuer Computeranwendungen, ICA-1, Pfaffenwaldring 27, 70569, Stuttgart, Germany.
	 (2) Departamento de F\'{\i}sica, Universidade Federal do Cear\'a, 60455-970, Fortaleza, CE, Brazil.}
\end{center}

\begin{abstract}
We present a simple one-dimensional model to describe the evolution of a transverse dune field. The model is characterized by the distances between the dunes and their heights, which determine the inter-dune sand flux. The model reproduces the observation that the dunes in a given field have approximately all the same height. We find that this result is independent on the initial configuration of the field, as well as on coalescence effects between migrating dunes. The time for the final state to be reached is studied as a function of the relevant phenomenological parameters.

\end{abstract}

\end{frontmatter}

\section{Introduction}
There are over hundred different types of dunes that have been classified by geomorphologists according to their shape and size \cite{lancaster}. Dunes can be found in deserts, in coastal areas, on the sea-bottom \cite{berne_1989}, and even on Mars \cite{thomas_1999}. Although no general theoretical model has been developed up to now to explain this wide variety of dune shapes, it is well known that the dune morphology is mainly determined by the amount of available sand and by the behavior of the wind over the year. The simplest type of dunes occurs when there is not very much sand, and when the direction of the blowing wind doesn't change with time. These dunes have a crescent-like shape and are called ``barchans''. Since the pioneering work of Bagnold in 1941 \cite{bagnold}, several field measurements led to a good understanding of barchan dunes. In particular, it has been observed that these dunes move without changing their shape with a velocity which is inversely proportional to their height \cite{kroy_2001}. Fields with hundreds of barchans are found for instance in Peru \cite{finkel_1959,hastenrath_1967,lettau_1969}, Namibia \cite{slattery_1990} and Marroco \cite{oulehri_1992,sauermann_2000}. In a recent work \cite{lima_2002}, a two-dimensional model to describe the collective motion of barchan dunes has been presented, which was able to reproduce the same kind of spatial correlations between the dunes as in real dune fields.

As the sand availability becomes larger, parallel dunes perpendicular to the wind direction, which are called ``transverse dunes'', arise for a wind blowing steadily from the same direction. Ensembles of transverse dunes, with crest to crest distances ranging from a few meters to over 3 km, are found to cover about 40\% of all terrestrial sand seas, being common in the Northern Hemisphere \cite{tsoar_2002,burkinshaw_1993}, and also dominating the sand landscape on Mars \cite{thomas_1999}. The last few years have seen an increasing interest in the investigation of the geomorphological aspects and the evolution of transverse dunes \cite{werner_1999,momiji_2000,warren_2000}. In spite of this effort, very little is known about these dunes, and several phenomena are still not well understood, as for example the effect of coalescence between dunes with different velocities \cite{besler}. 

This work reports on a simple one-dimensional model for the evolution of transverse dunes in a field, which reproduces the observation that these dunes have generally the same height and velocity of propagation. This is independent on the initial configuration of the field and on coalescence effects between close dunes with different velocities. In our model each dune is considered to have zero width and infinite length, but a variable height, which determines the associated propagation velocity and that also controls the interaction between neighbouring dunes. The paper is organized as follows. In section 2, the model is presented; the results for both cases with and without coalescence are discussed in section 3; conclusions are made in section 4. 

\section{The Model}
We consider a set of $N$ dunes distributed along the $x$ axis, each one with a variable height $h_i$, but zero width. The dunes are allowed to move in the $x$ direction with a velocity $v_i \equiv dx_i/dt$ given by:
\begin{equation}
v_i = \frac{a}{h_i} \label{eq:v},
\end{equation}
where $x_i$ is the position of the $i-$th dune, $i=1,2,\ldots,N$, and $a$ is a constant phenomenological parameter that contains information about the wind speed, the grain size, etc.

The behavior of the sand flux on a dune field is crucial to the evolution of the field dynamics, as interactions between dunes strongly depend on the sand transport among them. For a dune that migrates shape invariantly through sand avalanches on its slip face, the sand flux $\phi$ over the dune behaves as $d{\phi}/dh = $constant, where $h(x)$ is the height profile of the dune. For the model studied in this paper, we consider the height $h_i$ of the dune at the position of the crest, and set for the $i-$th dune:
\begin{equation}
{\phi}_i = b\,h_i \label{eq:phi},
\end{equation}
where $b$ is our second phenomenological parameter, again depending on wind speed, grain size, etc. Note that equation (\ref{eq:phi}) means that the rate of sand transport over the crest is larger for the bigger dunes than for the smaller ones. As a consequence, exchange of sand is expected between neighbouring dunes with different heights, as a bigger dune tends to loose sand more rapidly than a smaller one.

To model the inter-dune interaction, we first observe that for a two-dimensional slice of a dune, the mass of sand in the dune, or equivalently, its area $A$, increases with the height of the dune as $A \sim h^2$. On the other hand, the amount of sand transported depends on the difference between the sand fluxes on the crests of the neighbouring dunes, i.e. if $dA_i$ is the volume change for the $i-$th dune during a time interval $dt$, then $dA_i \sim (dh_i)^2 \sim ({\phi}_i - {\phi}_{i-1})$. Thus, we write:
\begin{equation}
\frac{dh_i}{dt} = -c \times {\sqrt{|{\phi}_i-{\phi_{i-1}}|}} \times {{\mbox{sign}}({\phi}_i-{\phi_{i-1}})} \label{eq:dh/dt},
\end{equation}
where $c$ is the third parameter of our model that depends on the shape of the dune, and the last factor on the right side ensures that more sand is transported from the larger dune to the smaller one. As in a real transverse dune field, we set the sand influx to point in the positive wind ($x$) direction and assume it to be constant. We call the sand influx ${\phi}_0$, which can be seen as an effective dune fixed at the position $x_0=0$, with a constant height $h_0={\phi}_0/b$. Equations (\ref{eq:v}), (\ref{eq:phi}) and (\ref{eq:dh/dt}) are repeatedly iterated, which gives the time evolution of the dune field. 

\section{Results and discussion}
We studied the evolution of a set of $N$ dunes with initial positions $x_i(0)$ and heights $h_i(0)$, for different configurations with $N=10$, as shown in Figure 1. In the first group of simulations (I), the dunes are equally spaced along the $x$ axis at $t=0$, and have the same heights $h_i(0) = H, i=1,2,\ldots,N$ (Figure 1(a)). In the second group (II), we take random values for the heights $h_i(0)$ chosen from a homogeneous distribution between 0 and 1, but still the dune spacing is regular (Figure 1(b)). And in the third group (III), both the values for the heights and the positions of the dunes along the $x$ axis are chosen randomly (Figure 1(c)). 

The trajectory of the dunes along the $x$ axis and the changes in their heights are shown as a function of the time in Figures 2(a) and 2(b) respectively, for particular realizations corresponding to each one of the initial states I, II and III. The values of the phenomenological parameters are $a=300{\mbox{m}}^2/{\mbox{year}}$, $b=1500\,{\mbox{year}}^{-1}$ and $c=0.05$, and the sand influx is chosen to be ${\phi}_0 = 2.25 \times 10^4\,$m/year $\approx 4.3\,$cm/min. In the simulations, each iteration corresponds to $0.01$ year. As can be seen from these figures, after a certain number of iterations $t \sim T_{\infty}$, all dunes reach the same height $h_i(T_{\infty}) = {\phi}_0/b = 15$m, and consequently the same velocity $v_i(T_{\infty}) = a/h({T_{\infty}}) = {(a \times b)}/{\phi}_0 = 20$m/year, independently on the initial configuration. 

In the discussion above, the distances between the dunes were large enough that their relative order was the same during the whole evolution of the field. However, if a smaller, faster migrating dune is very closely behind a larger, slower one, they can reach the same position after some finite time. Dune coalescence has been studied for instance in Ref. \cite{lima_2002}, in which a critical distance $d_{\mathrm{min}}$ between two neighbouring dunes was considered, below which they collide and become one single dune. In our one-dimensional model, we set $d_{\mathrm{min}} = 0$. For simplicity, dune coalescence is implemented here through the emergence of a single dune, with height $h_{\mathrm{c}} = \sqrt{{h}^{2}_{i} + {h}^{2}_{i+1}}$. Consequently, the number of dunes $N$ in the field decreases by unity, whenever two dunes coalesce.

The main plot of Figure 3 shows the variable $x_i(t)$ for a set of $N(0)=20$ dunes with random values of $x_i(0)$ and $h_i(0)$ for $t=0$. The values of ${\phi}_0$ and of the parameters $a$, $b$ and $c$ are the same as in Figures 2(a) and 2(b), but now their distances in the initial state are much smaller. The number of dunes as a function of time for the realization in this figure is shown in the upper-left-hand-corner inset. Note that this number decreases monotonically until the last coalescence between dunes has occured, and the final number of dunes, $N(\infty)=3$ has been reached. As we can see from the bottom-right-hand-corner inset, coalescence of dunes doesn't modify the result that the dunes in the field do all have the same height in the final state. 

\begin{figure}
\begin{center}
\includegraphics*[width=.9\columnwidth]{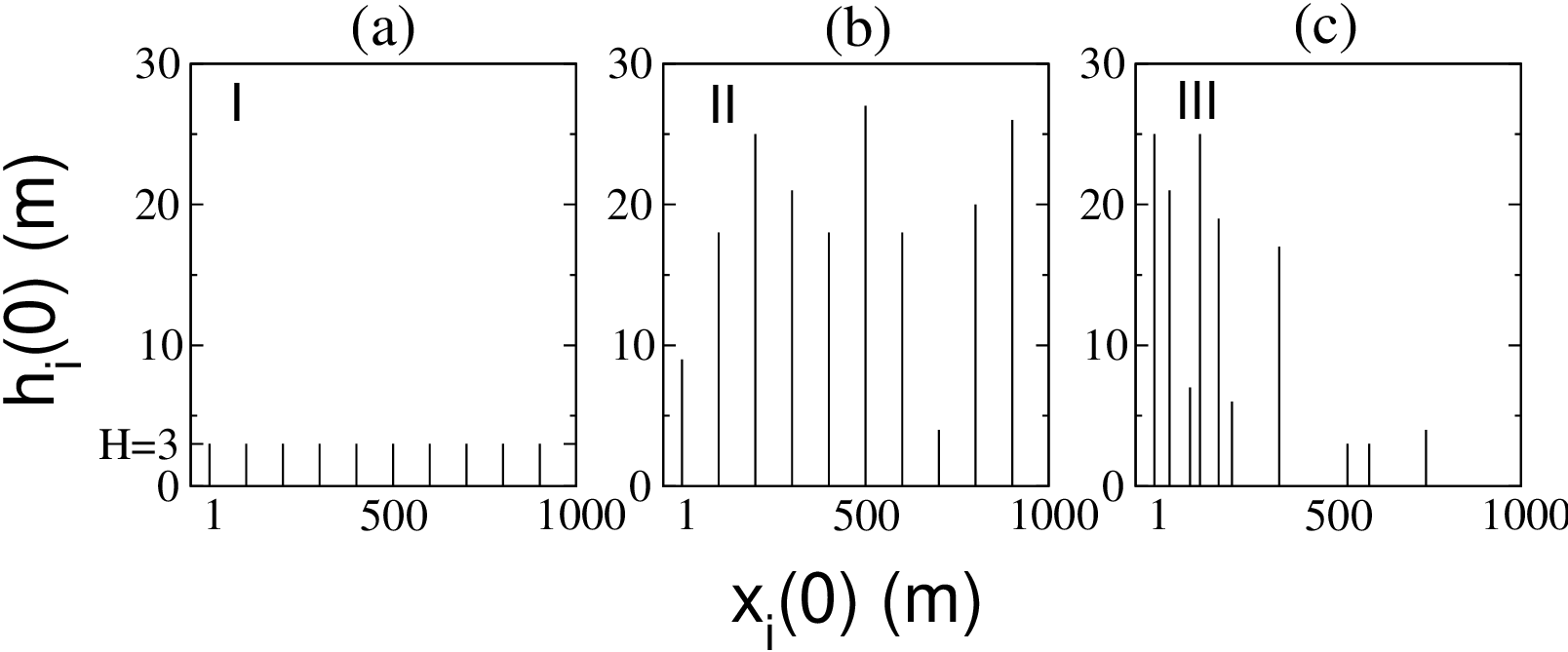}
\caption{Different realizations of a set of $N$ dunes with positions $x_i(0)$ and heights $h_i(0)$ at $t=0$. In (a) the dunes are equally spaced, and have the same height $H$ (group I). Dunes in group II still have regular spacing but random heights (b), while in group III, both the positions and heights are random (c).}
\label{fig:fig1}
\end{center}
\end{figure}
\begin{figure}
\begin{center}
\includegraphics*[width=.9\columnwidth]{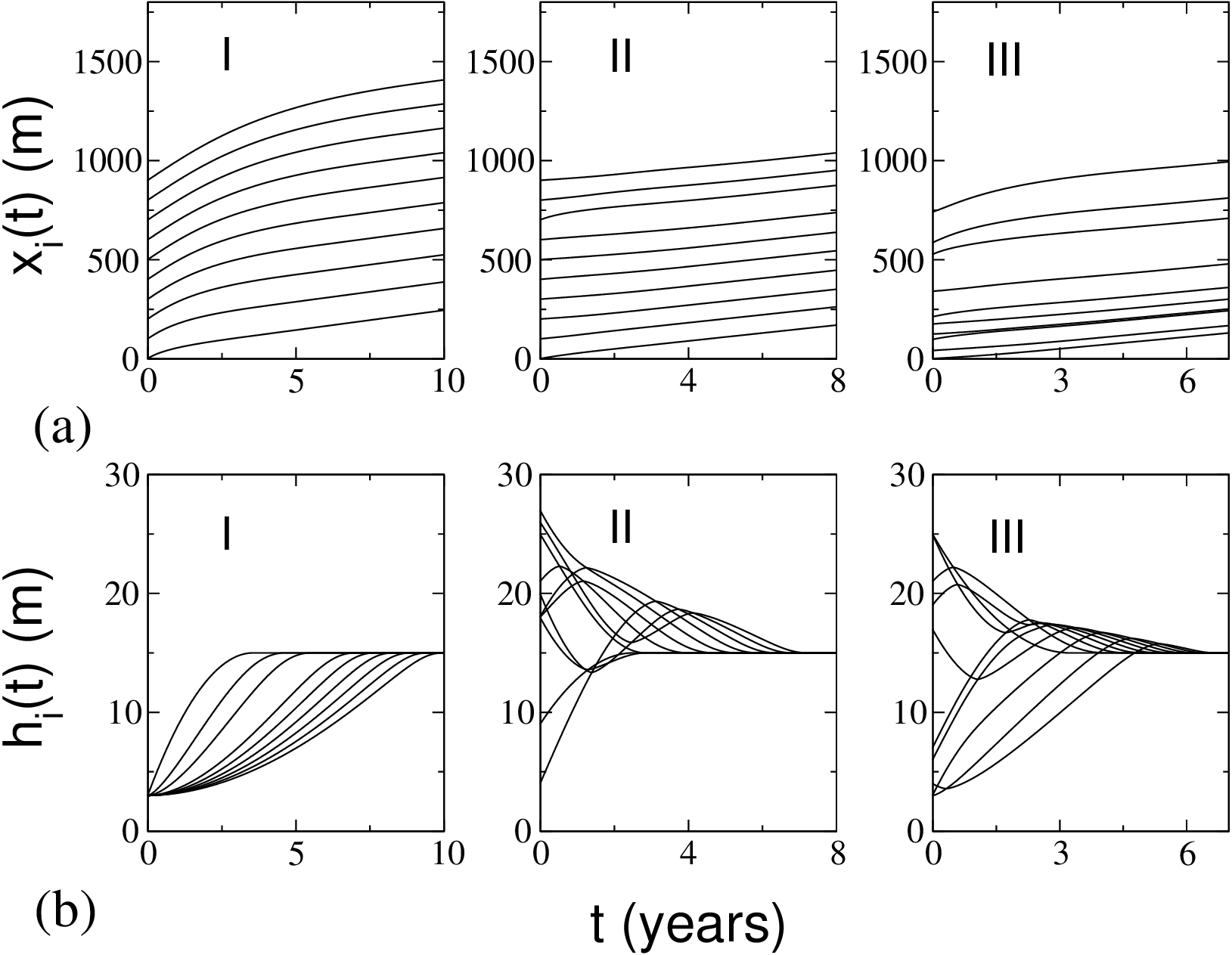}
\caption{(a) Trajectory of the $N=10$ dunes along the $x$-axis as function of time, for each of the realizations shown in Figure 1. The evolution of the heights of the dunes for each case is shown in (b). The values of the parameters used are $a=300\,{\mbox{m}}^2/{\mbox{year}}$, $b=1500\,{\mbox{year}}^{-1}$, $c=0.05$, and ${\phi}_0 = 2.25 \times 10^4$m/year. See text for details.}
\label{fig:fig2}
\end{center}
\end{figure}

\begin{figure}
\begin{center}
\includegraphics*[width=.9\columnwidth]{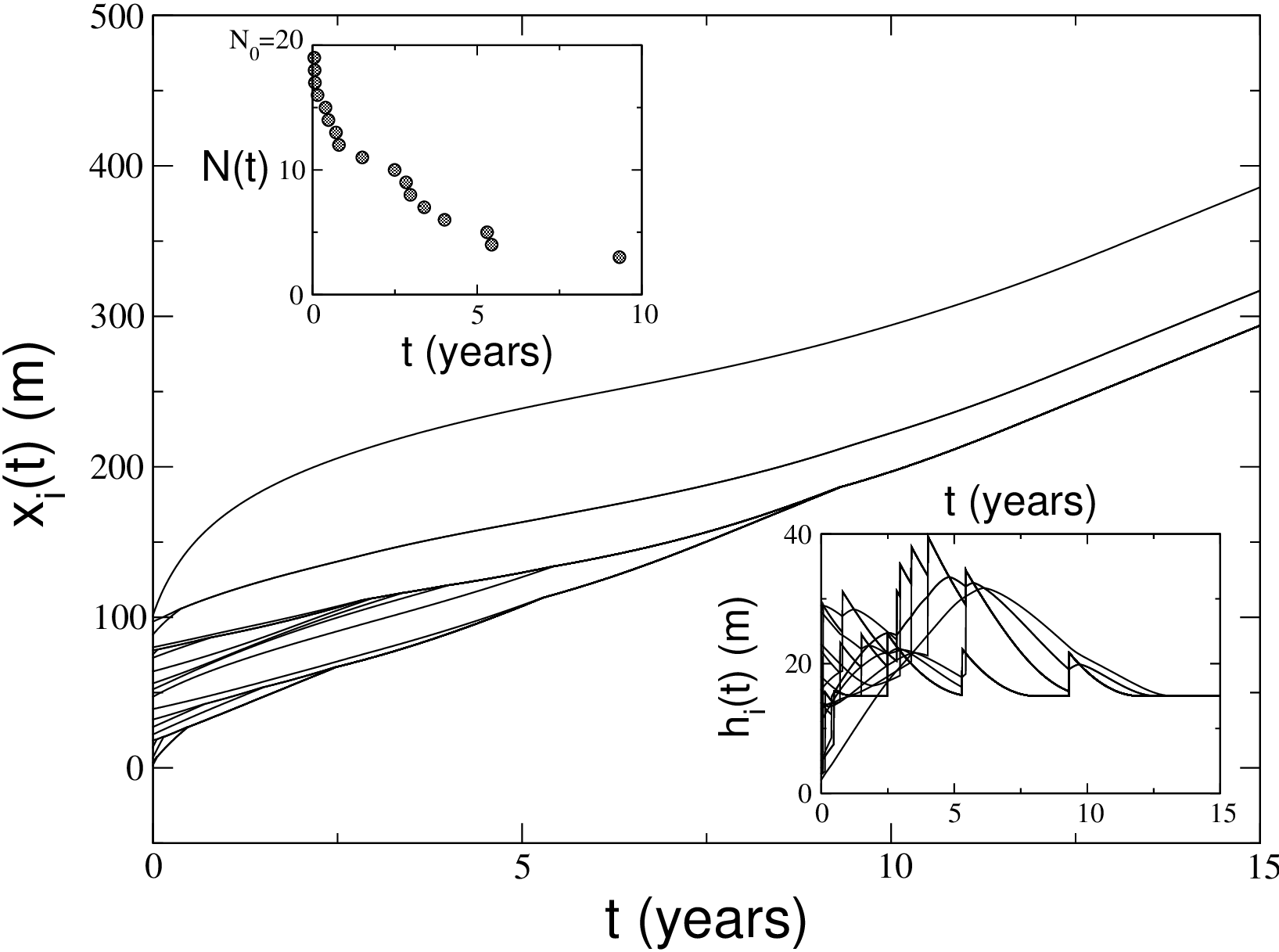}
\caption{The main plot shows the time evolution of the position $x_i(t)$ for a set of $N(0) = 20$ dunes with coalescence for the values of the parameters used in Figure 2. The number of dunes as a function of the time, $N(t)$, is shown in the upper-left-hand-corner inset of this figure. From the bottom-right-hand-corner inset, we can see that dune coalescence doesn't change the fact that all dunes have the same height for $t \rightarrow T_{\infty}$.}
\label{fig:fig3}
\end{center}
\end{figure}

For a given distribution of dunes in the field at $t=0$, the number $N(\infty)$ of dunes in the final state tends to be larger for smaller values of the parameter $a$, the other parameters $b$, $c$ and ${\phi}_0$ being fixed. This is because $a$ controls the velocity of the dunes, and coalescence is more frequent in a field with faster dunes. In the same way, the closer the dunes are in the initial state, the more they are going to coalesce on average. As an example, we show in Figure 4 the behavior of $N(\infty)$ as a function of the length $L$ of the system, i.e. the initial distance between the first dune and the last one in the field, for a set having a fixed number of $N(0)=50$ dunes, using the same values of the phenomenological parameters as in Figure 1. As we can see in this figure, the smaller (larger) the value of $L$, the more (less) the dunes tend to coalesce. The statistical error bars are obtained from different realizations of dunes with random heights and positions at time $t=0$. The full line shown in Figure 4 corresponds to the fit:
\begin{equation}
N(\infty) = N(0)\,[1-q*\exp(-L/L^*)] \label{eq:fit},
\end{equation}
where $N(0) = 50$, $q \simeq 1$, and $L^{*} \simeq 1.34 \times 10^3\,$m. 
\begin{figure}
\begin{center}
\includegraphics*[width=.7\columnwidth]{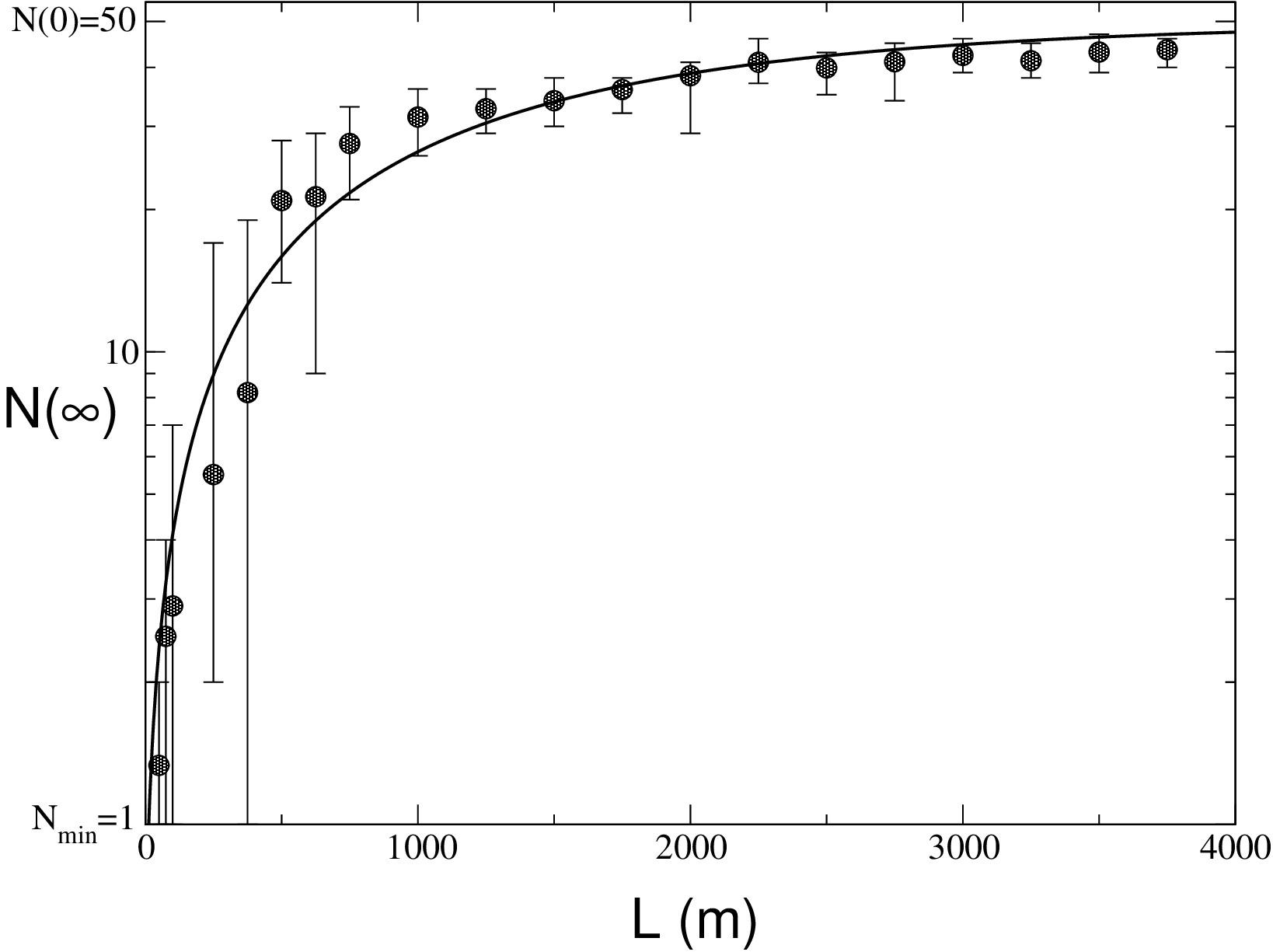}
\caption{Average number of dunes $N(\infty)$ for $t=T_{\infty}$, as function of the size $L$ of the system, i.e. the distance between the first dune and the last one in the initial state. The statistical error bars are obtained from an average over different realizations $(x_i(0), h_i(0))$ of $N(0) = 50$ dunes at $t=0$. The full line corresponds to the fit using the equation $N(\infty) = N(0)\,[1-q*\exp(-L/L^*)]$, with $q \simeq 1$ and $L^* \simeq 1.34 \times 10^3\,$m.}
\label{fig:fig4}
\end{center}
\end{figure}
In Figure 5, we show for three different values of $L$ the corresponding curves $N_L(t)$, averaged over five realizations of $(x_i(0), h_i(0))$, as a function of time $t$. The values of $L$ shown are $L=100$m (circles), $L=500$m (triangles) and $L=1000$m (squares). The dashed line corresponds to a straight line in a linear-log fit to the curve corresponding to $L=100$m. We can observe that $N_L(t)$ decays approximately logaritmically with time, independently on $L$. After a variable time interval $T_c \leq T_{\infty}$, coalescence stops and an asymptotic value $N(\infty)$ is reached. 
\begin{figure}
\begin{center}
\includegraphics*[width=.7\columnwidth]{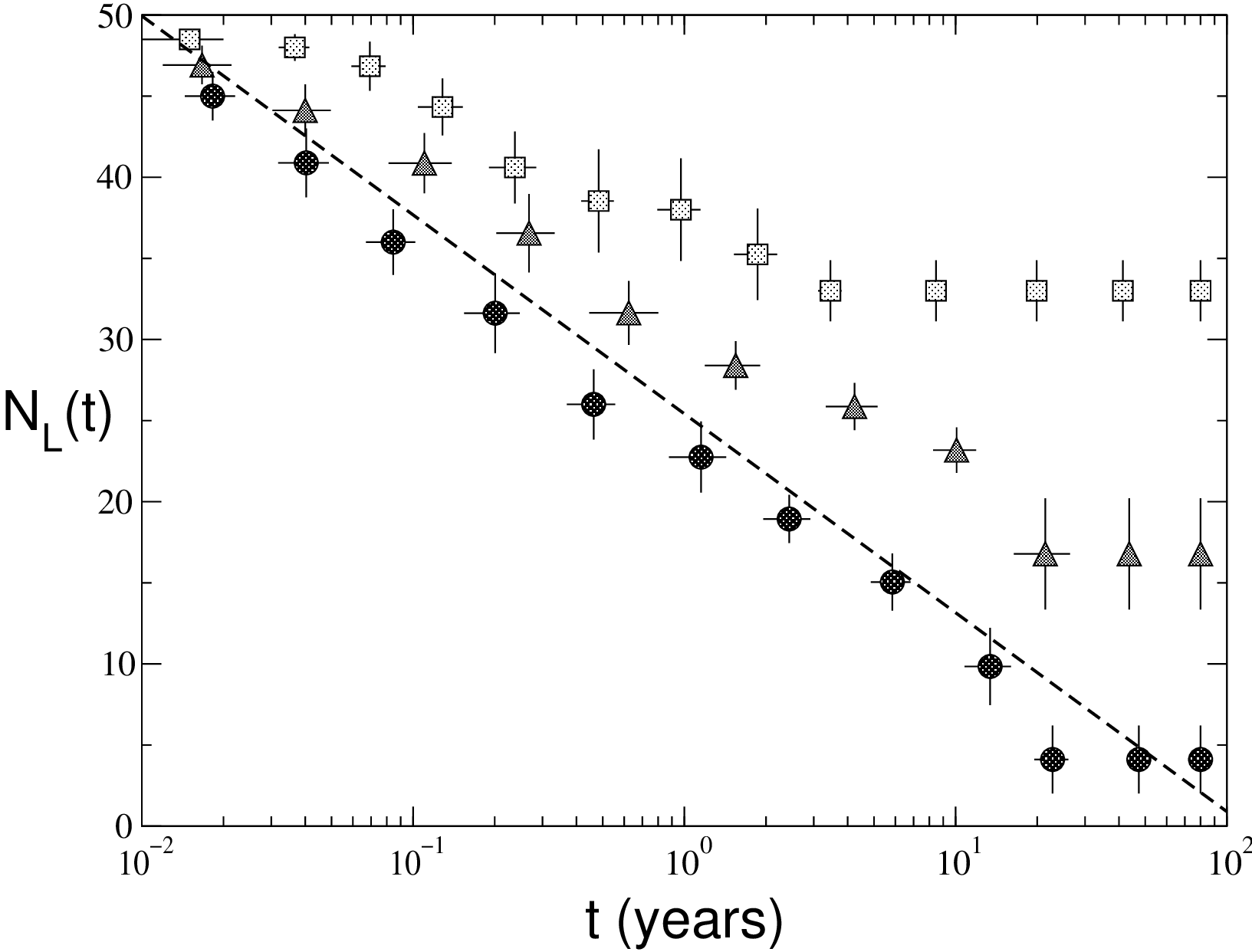}
\caption{Three $N_L(t) \times t$ curves corresponding to averages over different initial realizations of the type III $(x_i(0),h_i(0))$ of a set of $N(0)=50$ dunes, which are let to evolve with coalescence. The curves correspond to $L=100$m (circles) $L=500$m (triangles) and $L=1000$m (squares), where $L$ is the distance between the first dune and the last one in the initial state of the field. Time is plotted on a logarithmic scale. The dashed straight line is a guide to the eye meant to fit to the curve for $L=100$m (circles).}
\label{fig:fig5}
\end{center}
\end{figure}

As all dunes attain the same height $h_0 \sim {\phi}_0$ for $t = T_{\infty}$, the time to reach the final state depends on the strength of the sand flux in the field, ${\phi}_0$. A plot of $T_{\infty}$ as a function of ${\phi}_0$ for the initial configuration I of Figure 1a is presented in Figure 6. As we can see from this figure, $T_{\infty}$ scales after a transient interval of variation in ${\phi}_0-bh_i(0)$:
\begin{equation}
T_{\infty} \sim \sqrt{{|{\phi}_0-{\phi}|}} \label{eq:Txphi},
\end{equation} 
i.e. the closer the value of ${\phi}_0$ to $\phi = bh_i(0) = bH$, the more rapidly the system will reach the final state. From equation ({\ref{eq:dh/dt}}), we see that the phenomenological parameter $c$ controls the rate of sand transport between neighbouring dunes. We have found that the number of iterations to reach the final state decreases as $1/c$, which in fact is independent on the initial height $H$ of the dunes.  
\begin{figure}
\begin{center}
\includegraphics*[width=.7\columnwidth]{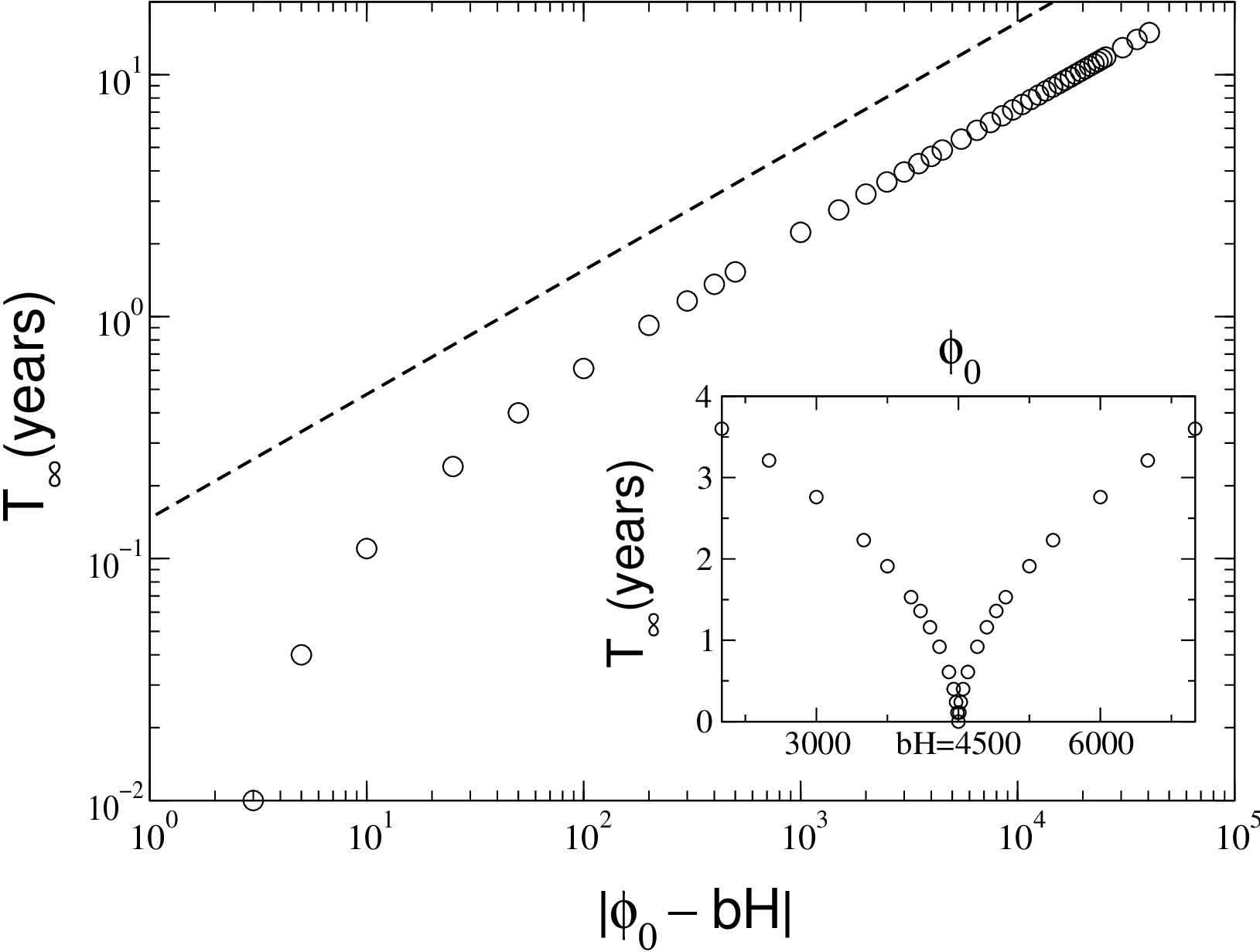}
\caption{Dependence of the time $T_{\infty}$ to reach the final state, on the relative sand flux $|{\phi}_0 - bH|$ for the initial configuration I shown in Figure 1(a). The circles shown in the main plot correspond to the case ${\phi}_0 > bH$, and line represents equation (\ref{eq:Txphi}). The inset shows $T_{\infty}$ as a function of ${\phi}_0$ for the same set of dunes.}
\label{fig:fig6}
\end{center}
\end{figure}

\section{Conclusions}
We presented a simple one-dimensional model for a transverse dune field which reproduces the observation that the dunes in a field attain approximately the same heights and velocities. The fact that our results do not depend on the initial configuration of the field and on the details of coalescence effects between dunes is of particular interest, since the genesis of dune fields is still not understood by geographers and geologists. Of course, serious limitations emerge from considering dunes as points in a line. Further work is required to compare the results presented here with real dune field observations. 

\begin{ack}
We acknowledge V. Schwaemmle for his contribution in the initial stage of this work. We acknowledge A. O. Sousa for useful comments. E. J. R. Parteli acknowledges support from CAPES - Bras\'{\i}lia/Brazil. 
\end{ack}

\end{document}